# The impact of the AGN and the torus properties on the evolution of spiral galaxies

M. A. Abdulrahman,[1]★ K. A. K. Gadallah,[2] A. Ahmed[1] and M. S. Elnawawy[1]

[1]*Astronomy, Space science and Meteorology Department, Faculty of Science, Cairo University, Egypt*
[2]*Astronomy and Meteorology Department, Faculty of Science, Al-Azhar University, Nasr city, 11884, Cairo, Egypt*

**ABSTRACT**
For spiral galaxies, the active galactic nucleus (AGN) and some physical parameters that concern the host galaxy such as spiral arm radius and density can play an important role in the morphological evolution of these galaxies. Considering the gravitational effect of the central black hole as a feeding mechanism, the gas flows from spiral arms to the accretion disk. Accordingly, we constructed our approach and derived an equation for the AGN luminosity that depends on parameters such as the black hole mass and the spiral arm density. The galaxy samples were taken from a catalog of type 1 AGN from SDSS-DR7. In our model, we present the relation between the AGN luminosity and the black hole mass depending on the above physical parameters. We also investigated the relation between the black hole mass and the star formation rate for the galaxy sample. The physical properties of the torus, such as the spiral arm radius, density, the torus length, and the gas mass, and the star formation rate were explained in terms of the variation of the AGN luminosity. These properties are more effective in the evolutionary scenario of the spiral galaxy. Relative to the variation of the AGN luminosity, the evolutionary track is different based quantitatively on the star formation rate. In which the variation in the star formation rate is positively correlated with the AGN luminosity.

## 1 INTRODUCTION

The first active galactic nuclei (AGN) in nearby galaxies were observed and described by Seyfert (1943) who found spiral galaxies having stronger emission lines from their nucleus than usual and was named Seyfert galaxies. Later observations have revealed many types of AGN which exhibit different features such as seyfert type-1 and seyfert type-2 (hereafter, type 1 and type 2, respectively). Based on their optical emission lines, type 1 shows broad emission lines and type 2 shows only narrow emission lines (Khachikian & Weedman 1974). The standard unified model introduced by Antonucci (1993) defines the AGN as a central black hole (BH) surrounded by an accretion disk and dusty torus.

The AGN luminosity should be controlled by its evolution which depends on the feeding mechanism and available matter for feeding. Concerning the feeding mechanism, Alonso, Coldwell & Lambas (2014) found that barred active galaxies show an excess of nuclear activity more than unbarred ones which refers to the importance of bars in gas inflow to central regions. And Dubois et al. (2015) showed in their simulation that supernova (SN) feedback could alter the evolution of central BH during galaxy formation, where in strong SN feedback the energy released can destruct the dense clumps in the galaxy core preventing the evolution of BH due to the lack of cold gas. Concerning the available matter, Koss et al. (2021) found that AGN in massive galaxies ($>10^{10.5}$ M$_\odot$) tends to have larger molecular gas and gas fraction than inactive galaxies. And Franceschini et al. (1999) found that the histories of the BH accretion rate and stellar formation in host galaxies are similar. Also Heckman et al. (2004) studied the accretion-driven growth of BH at low redshift using type 2 AGN and found that the BH of masses less than $10^8$ M$_\odot$ that reside in moderately massive galaxies have accretion rate time scale that is comparable to the age of universe.

Since AGN is powered by a central BH which resides at a host galaxy of certain physical properties such as stellar mass and star formation rate, the evolution of AGN should be related to its host galaxy. In the following we will present some previous works on studying the relation between AGN and host galaxy either by observational evidence or hydrodynamical simulations.

The accretion process caused by the central BH, results in a nuclear activity that appears as a feedback on the ambient medium. This activity has different scenarios such as winds, radiation pressure, and jets (Zubovas & King 2014a; Fabian, Vasudevan & Gandhi 2008). The model made by King & Pounds (2003) explained the energy and momentum of large scale outflows. According to this model, the AGN radiation pressure can launch a relativistic wind from very close in, where outflow can emerge from a photosphere of a radius a few tens of schwarzchild radius ($R_s$) given by $R_s = \frac{2GM}{c^2}$, where $M$ is the BH mass. This wind shocks against the ambient medium producing an outflow. When the BH reaches a critical mass given by Zubovas & King (2014b),

$$M_\sigma \approx 3.67 \times 10^8 \sigma_{200}^4 \text{ M}_\odot,$$

where $\sigma$ is the velocity dispersion in host galaxy spheroid, the AGN outflows become energy driven and cannot be cooled any more which can cause gas loss and consequently affects the star formation in the host galaxy. Also Wagner et al. (2016) showed that the feedback can be positive such as triggering star formation by using its energy in pressure triggered collapse or negative such as quenching star formation by the loss of material from the host galaxy.

★ E-mail: mohamedabdulazez9@gmail.com

On the other hand, time evolution simulations of the impact of the jet emerging from AGN on the ambient interstellar medium (ISM) made by Wagner, Bicknell & Umemura (2012) showed that the effect of the jet depends on the size of the ambient clouds. For large clouds (>50 *pc*), it increases the star formation rate, but for small clouds (∼10 *pc*), it causes explosion of cloud and quenching of star formation. So negative and positive feedback can co-exist depending on the density of the ambient ISM. In the simulations made by Dubois et al. (2013), the jet power can be considered as a reason of the material loss which can transform the disk galaxy into red elliptical galaxies by the quenching of star formation. Also Mukherjee et al. (2018) performed simulations to study the effect of a relativistic jet on the ambient gaseous disk and found that, depending on jet power, the ISM density, and jet orientation, the star formation can be enhanced or quenched. These authors found that jet can contribute in increasing the velocity dispersion of ambient ISM. Accordingly, the effect of AGN on its host galaxy is relevant to the host galaxy properties.

Observational studies made by (Ferrarese & Merritt 2000; Gebhardt et al. 2000) revealed a correlation between the BH mass and the velocity dispersion of the host bulge. Reines & Volonteri (2015) found a correlation between the BH mass and the stellar mass of the host galaxy in which the BH mass increases with increasing the stellar mass of its host galaxy. This is consistent with the observations made by Bilata-Woldeyes et al. (2020) using data from BAT-SWIFT to study the relation between morphology of the host galaxy and the AGN properties such as the Eddington ratio and BH mass. In these observations, the BH masses are larger in elliptical galaxies than in spiral galaxies.

Recent studies made by Dittmann & Miller (2020) investigated the growth of the central BH by assuming a merging scenario with the compact objects formed in the accretion disk. The study by Tartenas & Zubovas (2019) showed the feeding of the AGN by the dynamical perturbations. It also showed that a collision between circumnuclear ring and molecular cloud that can be an efficient fueling mechanism, depending on the angle of collision. So the fueling mechanism has a crucial role in driving the relationship between the AGN and its host galaxy. As a way to further examine how such a relation between the AGN and host galaxy exists, Smethurst et al. (2016) studied the star formation history of type 2 AGN and inactive galaxies. In this study, the gas reservoir in the host galaxy is the main source for BH fueling as first examined and mentioned by Magorrian et al. (1998), where both of nuclear activity and star formation are related to the host galaxy gas reservoir.

To further understand this mutual effect between the AGN and the host galaxy, Valentini et al. (2020) performed a simulation of the galaxy disk to study gas accretion models. This simulation showed how the AGN feedback on the multiphase of the ISM of hot and cold phases can affect the co-evolution of the BH and its host galaxy and found that the accretion of cold gas is more effective in BH growth rate than hot gas, and the gas accretion contributes in the BH growth more than mergers with other BHs.

So in this paper we study the evolution of isolated spiral galaxies in terms of AGN contribution by investigating the effect of AGN on the spiral arms of spiral galaxies and how the spiral arms can control the AGN activity as being a source of fueling of the central BH. Depending on the fueling of AGN by driving the gas from the spiral arm under the gravitational force of the central BH to derive an Equation for the AGN luminosity that depends on parameters such as the BH mass and the spiral arm radius which is a measure of the amount of gas in it. We also investigate the BH mass-AGN luminosity relation and how it can be affected by changing physical parameters such as the spiral arm radius and the torus length.

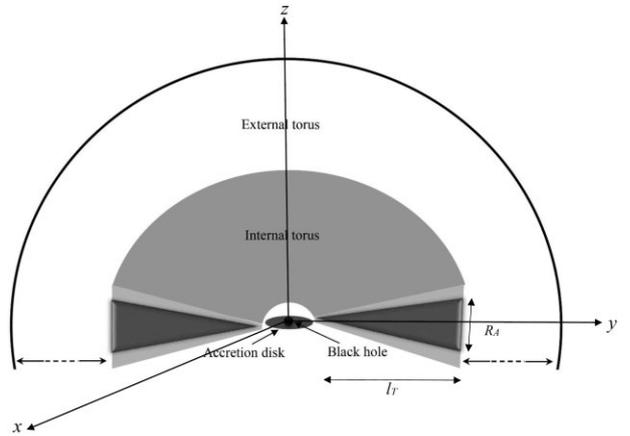

**Figure 1.** Schematic chart of a simple spiral galaxy having two arms.

We hope to construct a physical scenario for how the AGN-galaxy co-evolution works. This paper is presented as follows. The model approach is explained in Sections 2 and the data of the galaxy sample is given in Section 3. Results and the discussions are presented in Sections 4 and 5, respectively, while the conclusion is given in Section 6. In this work, the cosmological parameters were taken with $H_0 = 70.0$ km s$^{-1}$ Mpc$^{-1}$, $\Omega_m = 0.30$, $\Omega_\lambda = 0.70$.

## 2 MODEL APPROACH

For a spiral galaxy having a simple galactic disk with two arms, we assume that most of the mass of the galactic matter is concentrated within these dense arms while the rest of the galactic medium of the host disk is very diffuse medium with very low density that can be neglected. Based on the simple morphology of the unified model, we assume that the torus is divided to internal and external sides around the the AGN as shown in the schematic chart in Fig. 1. In this chart, the internal torus is facing the AGN while the external torus represents the outer region of the galactic disk.

The internal torus region is assumed to partially contain these arms. In which the feedback of the BH activity is more efficient on the matter with these arms. In this context, Yu et al. (2022) found that the spiral arm is efficient in transporting the gas to central region. Accordingly, we assume that the material transfers in a conical shape from the closest region within the internal torus into the accretion disk through a spiral arm. This shape makes the spiral arm radius varies from smaller at the contact point with the accretion disk to larger radius within deeper internal region of the torus (neglecting the environmental effects on the galactic outskirts).

For a particle of mass ($m$) in the accretion disk at a distance ($r$) from the central BH, it moves with a velocity ($v$) where its angular momentum ($\Omega$) is given by,

$$\Omega = m_{AD} v_{AD} r \quad (1)$$

where $m_{AD}$ refers to the mass of accretion disk.

Due to the gravitational effect of the BH, the particles around the BH experience a torque pushing them inward into the BH. As the matter in the accretion disk loses angular momentum the matter spirals into the BH and this gives the chance to the material in internal torus to move to accretion disk.

So we can assume that the change of angular momentum equals the torque on the gas in the accretion disk. Accordingly, the change of *r* with time is a consequence of the change of the gravitational radius ($r_g$) of the BH with time which indicates the growth of BH or its evolution, hence,

$$\frac{dr}{dt} = \frac{dr_g}{dt} = \frac{\Gamma_{gas}}{m_{AD}v_{AD}} \quad (2)$$

where $\Gamma_{gas}$ is the torque on the gas in the accretion disk.

Considering a gaseous disk where the gravitational force is due to a central BH, the torque on gas at a radius (*r*), as given by Netzer (2013), is:

$$\Gamma_{gas}(r) = \dot{m}(GMr)^{1/2} f(r) \quad (3)$$

where, $\dot{m}$ is the radial mass inflow rate, *M* is the BH mass, and $f(r) = 1 - (r_{in}/r)^{1/2}$. The $r_{in}$ parameter represents the radius at which the torque on gas is zero and it falls in a non-circular orbit into the BH which is known as the inner-most stable circular orbit (ISCO).

Using $r_g = \frac{GM}{c^2}$ where *M*, *G* and *c* are the BH mass, the gravitational constant and the speed of light, respectively, we have

$$\frac{dr_g}{dt} = \frac{G}{c^2} \frac{dM}{dt} \quad (4)$$

It is known that the BH grows due to the accretion of matter with taking into account that not all matter is accreted. A fraction ($\delta$) of this matter does not accrete but it escapes due to the feedback of the AGN such as radiation pressure and wind from accretion disk (Zubovas & King 2014a) and the star formation taking place in the vicinity of BH (Dittmann & Miller 2020). So we can write the change of the BH mass with time as,

$$\frac{dM}{dt} = (1-\delta)\dot{M} \quad (5)$$

where $\dot{M}$ is the accretion rate given by $\dot{M} = \frac{L}{\eta c^2}$ while *L* is the AGN bolometric luminosity hereafter it is called $L_{AGN}$, and $\eta$ is the radiative efficiency.

Then equation (5) becomes,

$$\frac{dM}{dt} = (1-\delta)\frac{L}{\eta c^2} \quad (6)$$

From equations (2), (3), (4) and (6), the $L_{AGN}$ can be written as,

$$L_{AGN} = \frac{\eta c^4 \dot{m}(GMr)^{1/2} f(r)}{G(1-\delta)m_{AD}v_{AD}} \quad (7)$$

where a fast rotating BH has a spin parameter of 0.998, and $r_{in} = 1.24r_g$. It is possible to put $r = nr_g$ where *n* takes values of 20–40 indicating the mean disk size according to available data on variability of high redshift luminous AGN (Netzer 2013).

By considering the conical shape of the spiral arm facing the accretion disk, $m_{AD}$ can be written as,

$$\frac{dm_{AD}}{dt} = -\xi m_T \quad (8)$$

where $m_T$ is the mass of the internal torus, and $\xi$ is the transfer efficiency which could be thought as the fraction of medium clumpness (clumpy or smooth).

$$m_T = \frac{\pi}{3} R_A^2 \rho_T l_T \quad (9)$$

where $R_A$ is the spiral arm radius within the internal torus region characterized by a very dense conical shape. The parameters $\rho_T$ and $l_T$ are the density within the torus and its length, respectively. By assuming that the material is concentrated in the spiral arms at a certain time (t), for which, $\rho_T$ becomes mostly the arm density ($\rho_A$).

Substituting by (9) into (8), and integrating $\rho_A$ w.r.t time, the mass of accretion disk becomes

$$m_{AD} = \frac{\pi}{6} \xi R_A^2 \rho_A^2 l_T \quad (10)$$

Then the luminosity Equation in the present model becomes

$$L_{AGN} = 5.156 \times 10^{25} \frac{\eta n^{\frac{1}{2}} f(r) M \dot{m}}{v_{AD} \xi (1-\delta) l_T R_A^2 \rho_A^2} \text{ (erg s}^{-1}\text{)} \quad (11)$$

By setting $\delta = 0.1$, $\xi = 0.1$, $\eta = 0.1$, $n = 25$, $f(r) \sim 0.777$, $v_{AD} = 10^8 m/s$ as standard values, equation (11) can be re-written as

$$L_{AGN} = 2.792 \times 10^{18} \frac{M\dot{m}}{l_T R_A^2 \rho_A^2} \text{ (erg s}^{-1}\text{)} \quad (12)$$

Therefore, the effective parameters on the AGN luminosity are the mass inflow rate ($\dot{m}$), the internal torus length ($l_T$), the spiral arm radius ($R_A$), the spiral arm density ($\rho_A$), and the BH mass (*M*).

## 3 DATA

To compare with the calculations of equation (12) and in an attempt to check the validity of our BH mass-luminosity relation, a sample of the observational data was provided by a catalog of type 1 AGNs from SDSS-DR7 (Oh et al. 2015a). This catalog contains 5553 type 1 AGN with a redshift of $z \leq 0.2$. In which, galaxies were selected after applying some classification criteria for data of OSSY catalog. This catalog provides us with the logarithm of bolometric luminosity of AGN ($L_{bol}$) derived from a method developed by Heckman et al. (2004) using the luminosity of the [O III] $\lambda 5007$ emission line $L_{\text{O III}}$ as a tracer of nuclear activity where $L_{bol} \approx 3500\ L_{\text{O III}}$ (erg s$^{-1}$). It also provides us with the logarithm of the BH mass as derived by Greene & Ho (2005) depending on the luminosity and line width of the broad $H_\alpha$ line. The bolometric luminosity ranges from 42.09 to 46.77 erg s$^{-1}$ in logarithmic scale. Also the BH masses ranges from 6.13 to 9.29 $M_\odot$ in logarithmic scale.

As the galaxy selection is flux-limited, we have considered the Malmquist bias as a selecting criterion of our galaxy sample where the luminosity varies as a function of the redshift. In Fig. 2, we plot the bolometric luminosity across the redshift of 5553 galaxies as shown in the *top panel*, then we employ a flux-limit cut off with a minimum flux limit of $4 \times 10^{-12}\text{ erg s}^{-1}\text{cm}^{-2}$ to estimate the theoretical luminosity (solid line) using the formula of $L = 4\pi d_L F$ where $d_L$ is the luminosity distance. This luminosity distance was estimated according to the analytical approximation considering the case of flat cosmologies (Adachi & Kasai 2012). Accordingly, the number of galaxies in the sample has been reduced to 4954 galaxies as shown in the *bottom panel*. And the current range for the bolometric luminosity becomes from 42.88 to 46.77 erg s$^{-1}$ in logarithmic scale.

## 4 RESULTS

In this Section, we present the results that show the impact of the escaped fraction ($\delta$) and the clumpness fraction ($\xi$) mentioned above in equation (11) on the AGN luminosity. Then we used also equation (12) to show the effect of the spiral arm density ($\rho_A$) and the radius ($R_A$) on this luminosity. We showed the AGN luminosity versus the BH mass using the observational data along with those of equation (12). By using derived relations from previous observational works of (Domínguez Sánchez et al. 2012; Baron & Ménard 2019; Huang et al. 2012), we estimated the star formation rate, gas mass, and stellar mass of our sample the galaxies comparing the results from both of the observed data and equation (12). Finally, we used the results that we obtained in an attempt to put a suitable evolution scenario of the spiral galaxy in terms of AGN.

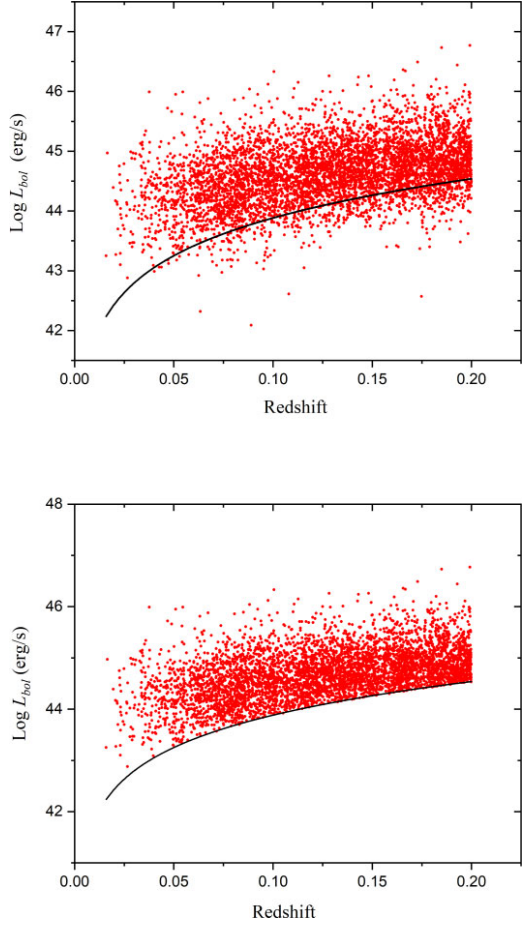

**Figure 2.** The bolometric luminosity of data sample (red dots) across redshift before (*top panel*) and after (*bottom panel*) applying the Malmquist bias where the solid line represents the flux-limit cut off of the galaxy sample.

### 4.1 The AGN luminosity versus torus properties

Using equation (11) with assuming $R_A = 100$ pc, $\rho_A = 10^{-10}$ kg m$^{-3}$, $l_T = 100$ pc, $\dot{m} = 0.1$ M$_\odot$ yr$^{-1}$, Fig. 3 shows how the AGN luminosity is affected by; the clumpness fraction of the medium ($\xi$) and the velocity of the material in the accretion disk ($v_{AD}$) in the top panels (*left and right, respectively*); the radiative efficiency ($\eta$) and the mean disk size (n) in the middle panels (*left and right, respectively*); and the escaped fraction ($\delta$) in the bottom panel. For each effective parameter, this luminosity was estimated with keeping the others constant. From this figure, the luminosity exponentially decreases with increasing both $\xi$ and $v_{AD}$, while it exponentially increases with increasing both of $\eta$, n, and in a logarithmic way with $\delta$.

In order to see how the BH mass affect the relation between the AGN luminosity and each physical parameter in equation (12), we present its behaviour at three different BH masses ($LogM = 6$, 7 and 8 M$_\odot$) as shown in Fig. 4 (*the assumed value for each parameter was chosen according to the best fit for the data sample shown later in Fig. 5*). From Fig. 4, the luminosity has a similar trends showing an exponential decrease with both of the $R_A$, $\rho_A$, and $l_T$ while it has differently an exponential increase with $\dot{m}$.

The spiral arm radius and density of a galaxy can be considered as a morphological parameters that demonstrate the host galaxy shape where the spiral arm contains most of the amount of material within torus. This matter is considered as the main feeding source for the central BH which controls the AGN luminosity. For which, the effect of both of spiral arm radius and density is shown in the top panels (left and right, respectively) of Fig. 4. Also Masoura et al. (2018) found that the AGN luminosity, in a host galaxy, depends on the position of this host galaxy from the main sequence line, depending on the available gas of the host galaxy. Since the accreted matter from the spiral arm travels through a path from spiral arm to the BH, the length of this path should also control the produced AGN luminosity. This path length is represented in our model by the internal torus length. The left top panel of Fig. 4 shows its correlation with the luminosity at three different BH masses, in which, the luminosity increases fast at smaller lengths but it decreases slowly at longer lengths.

All of the above physical parameters are related to the properties of the galaxy disk but the mass inflow rate or the accretion rate of the BH is related to the properties of the central BH. Its correlation with luminosity, shown in the left bottom panel Fig. 4, shows the remarkable increase of the luminosity with the mass inflow rate. This means that the luminosity doesn't depend only on the amount of the material flowing toward the BH but also the time taken by this material to reach the accretion disk.

For the data sample, the top panel in Fig. 5 shows the result with implying equation (12) and those of observed data fitted with a slope of $1.005 \pm 0.029$, for which, the standard deviation was adapted to $x = 1$, $y = 0.86$, giving a correlation coefficient of 0.49. To account for the dispersion of the scattered points, we calculated the residual which shows a normal Gaussian distribution as shown in the bottom panel.

Using the data sample, we present some physical properties concerning the host galaxy, such as the SFR, the stellar mass ($M_{stellar}$) and the gas mass ($M_{gas}$), to explain how these parameters can change across redshift. The observed data of our galaxy sample provides us with the $H_\alpha$ emission line luminosity, $L(H_\alpha)$, which we used to measure the SFR based on the SFR-$L(H_\alpha)$ dependence using equation (1) given by Domínguez Sánchez et al. (2012) where:

$$\text{SFR (M}_\odot \text{ yr}^{-1}) = 7.9 \times 10^{-42} \, L(\text{H}_\alpha) \, (\text{erg s}^{-1})$$

To estimate the stellar mass as a function of BH mass, we rearrange equation (9) given by Baron & Ménard (2019) to be as follows:

$$\text{Log}\left(\frac{M_{stellar}}{10^{11} \, \text{M}_\odot}\right) = \frac{Log(\frac{M_{BH}}{\text{M}_\odot}) - (7.88 \pm 0.13)}{(1.64 \pm 0.18)}$$

By assuming that stars form in molecular clouds, the gas mass can be measured using the HI 21 cm line. According to equation (1) given by Huang et al. (2012) using the stellar mass of $M_{stellar} > 10^9$ M$_\odot$ since our data range is 9.94–11.859 M$_\odot$ in log scale, the gas mass can be calculated in terms of the stellar masss as follows:

$$\text{Log}(M_{HI}) \approx 0.276 \, Log(M_{stellar})$$

For a deeper understanding of the relation shown in Fig. 5, we further investigated the SFR distribution along this relation. From Fig. 6, we can see that at a certain BH mass, the AGN luminosity decreases with decreasing the SFR. From equation (12), $\rho_A$ can be used as an indicator to the SFR. It is expected to have a clumpy medium with increasing the density in the arm. When star formation takes place, the ultraviolet radiation and the stellar wind injected energy, caused by massive stars. These could be destructive mechanisms to the molecular clouds leading to a decrease in the ISM density or cloud dispersion (Grudić et al. 2018; González-Samaniego & Vazquez-Semadeni 2020). Accordingly, the gas can then flow easily to the central region leading to high AGN luminosity.

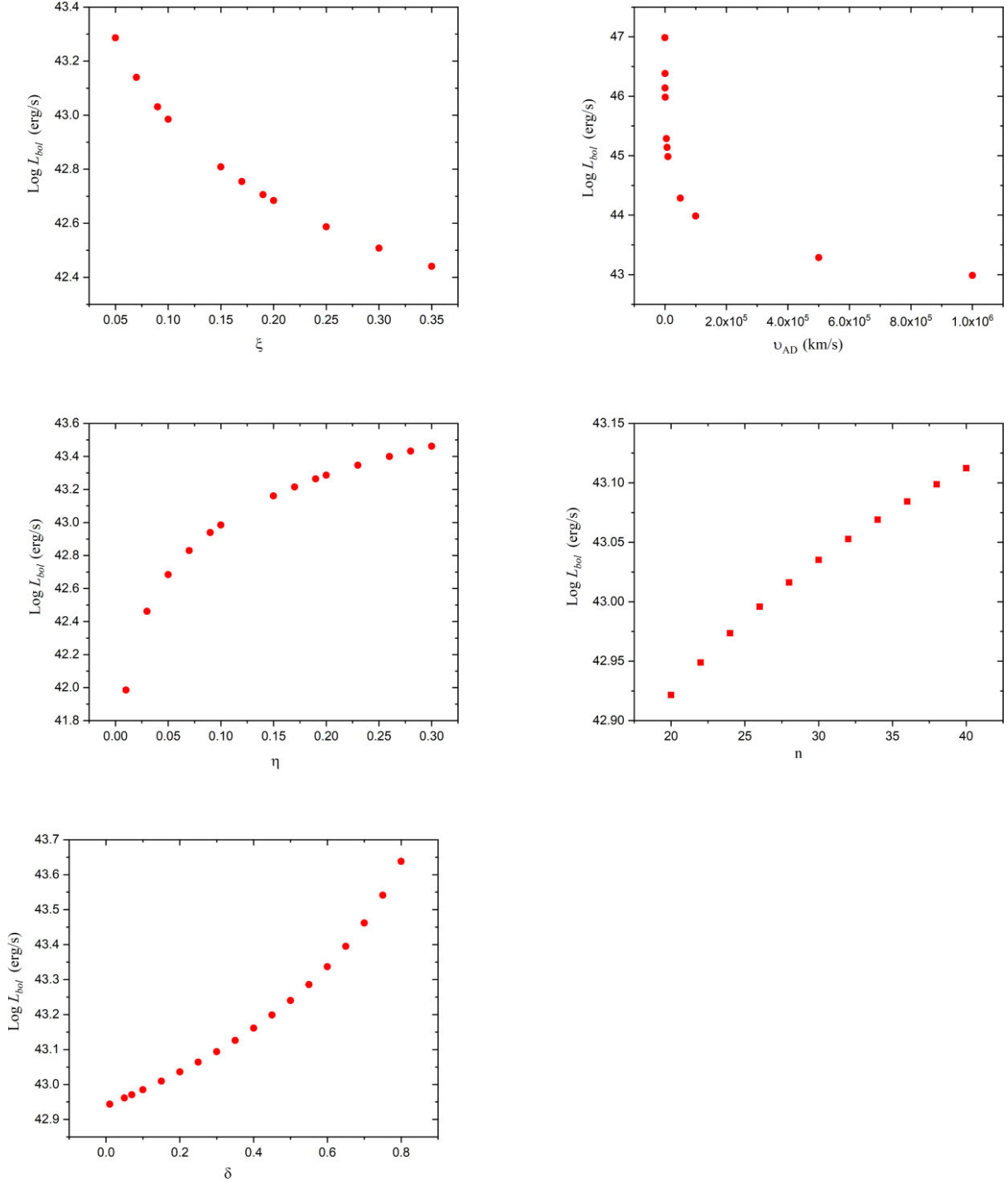

**Figure 3.** The AGN luminosity distribution as a function of the clumpness fraction and the velocity in the accretion disk (left and right top panels, respectively), the radiative efficiency and the mean disk size (left and right middle panels, respectively), and the escaped fraction from the BH accretion (bottom panel).

To implement the relation between SFR and AGN evolution, we present the SFR variation with BH mass as shown in Fig. 7. This figure shows a linear relation with a slope of 0.86 and a correlation coefficient of 0.49. It is obvious that this correlation fits well at smaller BH masses but SFR shows flattening at higher BH masses, deviating from the linear trend. Also from Figs 7 and 8 (bottom panel), we can see a variation in the SFR versus the BH mass and redshift, respectively. On average, the SFR is decreasing at low redshift values and low BH mass. This variation points to the relation between the BH mass and the SFR. In our work, we assume that the spiral arm is the reservoir of gas needed for the BH feeding and it can be seen from Fig. 7 along with Fig. 8 where gas in spiral arm is consumed by both of star formation and the BH feeding.

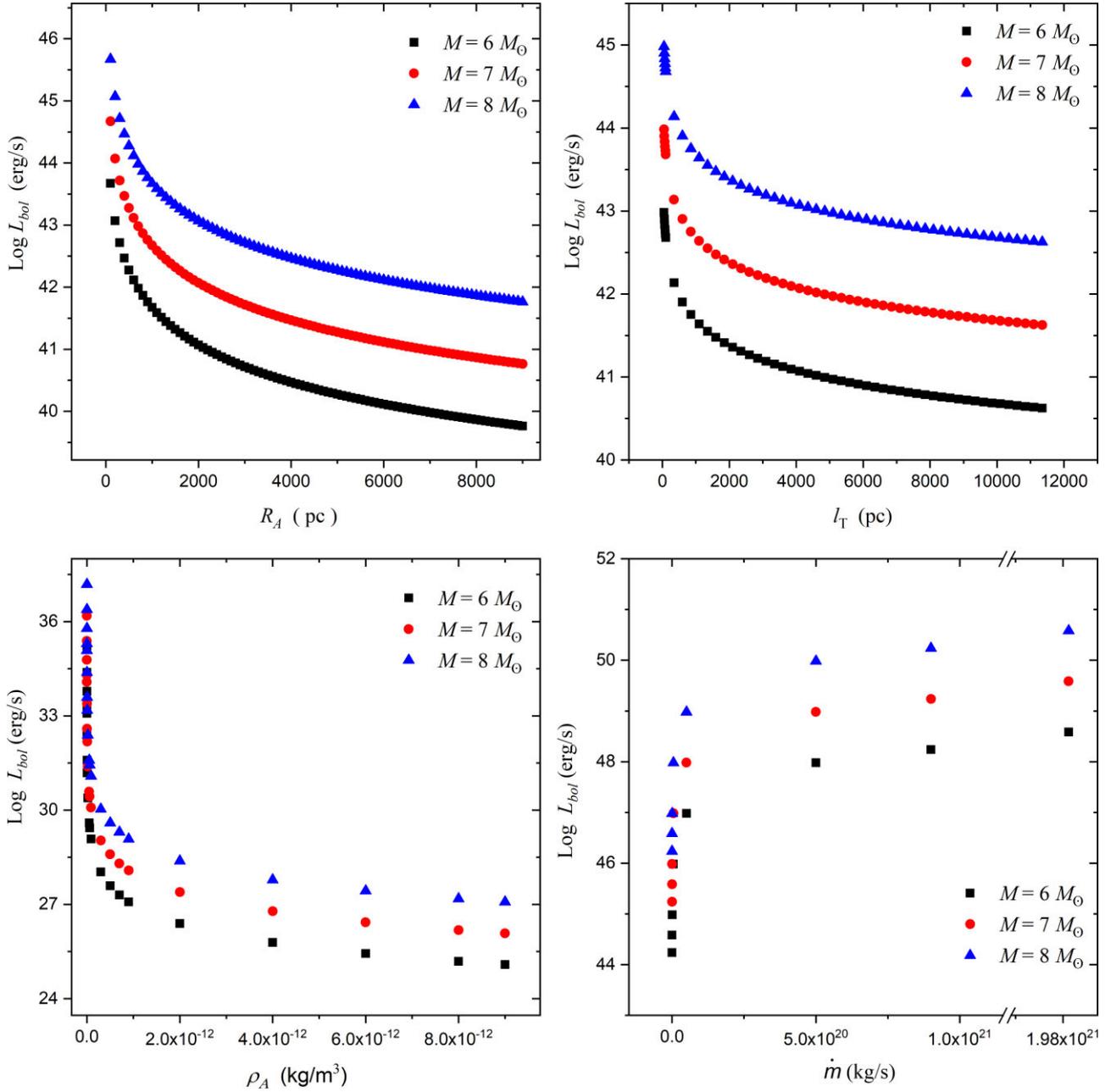

**Figure 4.** The bolometric luminosity of the AGN versus the spiral arm radius and the torus length (left and right top panels, respectively), and the spiral arm density and the accretion rate (left and right bottom panels, respectively). For these physical parameters, the results were estimated at 3 different values of BH mass of 6, 7, and 8 $M_\odot$ in log scale, assuming $R_A = 300\ pc$, $\rho_A = 10^{-12}$ kg m$^{-3}$, $l_T = 500\ pc$, $\dot{m} = 0.1$ $M_\odot$ yr$^{-1}$.

This gas consumption should decrease the gas content of the host galaxy (Scoville et al. 2016; Genzel et al. 2015). In Fig. 8 (top and middle panels), we can see that both of the gas mass and stellar mass are following the same trend with decreasing redshift. The decrease in each of them is slow and this can be used as evidence for the BH feedback such as jets and winds which could make the growth rate of the BH becomes slow. This can be used as evidence for the stellar cycle where the gas is converted into stars and then returns back through, for instance, the supernovae. Also previous work by Tacconi et al. (2008) showed that the $SFR/M_{gas}$ ratio is relatively constant.

### 4.2 Evolution of a spiral galaxy

We have showed that AGN luminosity is affected by the integration of all aforementioned physical parameters. Therefore, we tried to put an evolution scenario for an isolated spiral galaxy depending on these parameter and neglecting any merging and environmental effects.

For a spiral galaxy hosting a central BH, at first, the gas content of spiral arms is condensed to form clumpy clouds of gas as possible candidates for star formation to take place. If the gas gets condensed the amount of material available for the BH accretion decreases, hence the AGN luminosity decreases. This is why we get low AGN luminosity for high values of spiral arm radius and density. But due

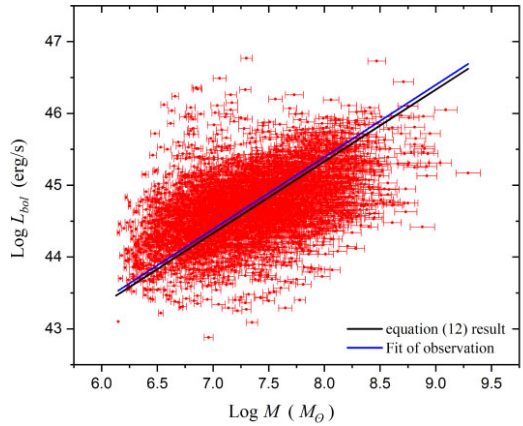

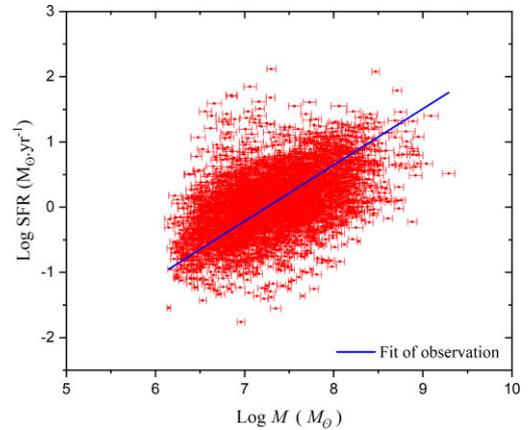

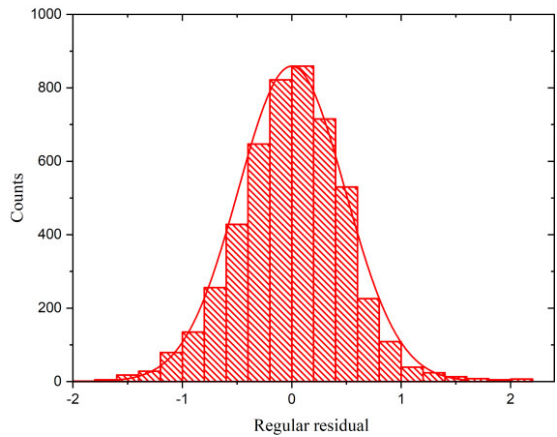

**Figure 7.** The star formation rate variation with BH mass.

to the star formation the luminosity emitted from the disk should be high.

As time goes and due to galaxy rotation and star formation, the gas in the spiral arms became dispersed which make it easy for the central BH to pull it producing high AGN luminosity. This is why we get high AGN luminosity for small values of spiral arm radius and density.

The accretion process has a time which is determined by the internal torus length and also by the accretion rate of the central BH. During the gas journey from the spiral arm to the central BH, it travels a certain path which is considered as the internal torus length. In our approach, we assumed that material is transferred through a conical path which can be representative of unbarred galaxy. This path length is in relevant to the galaxy size and also the existence of its bar which should alter the AGN luminosity. According to Alonso et al. (2014) who found that among their sample which includes barred and unbarred AGN, the barred galaxies exhibit a higher nuclear activity than unbarred ones. Also the length of the internal torus would affect the AGN luminosity and the activity time of AGN. Using the magnetic-hydrodynamical simulations, Rosas-Guevara et al. (2022) studied the evolution of barred massive disk galaxies. These authors found that barred galaxies have lower star formation rate and lower gas fraction compared to unbarred ones. This indicates that the existence or absence of a bar may increase or decrease the gas transport efficiency from the galactic disk to the accretion disk.

Linking the correlation between the AGN luminosity, gas mass, and SFR Shangguan et al. (2020), we can deduce the evolution scenario based on these physical properties of our galaxy sample. Using Figs 6 and 9 we can divide the evolution of a galaxy into three *phases*.

*Phase* 1 is the period of time where the gas is still condensed at spiral arms that we have a large gas mass galaxy at a certain SFR, the gas is consumed to form stars and there is not enough gas to be accreted by the central BH. For this period we should observe low AGN luminosity, low BH mass, and low stellar mass.

*Phase* 2 is the period of time where star formation has taken place that the stellar mass increases and the gas is dispersed whether by the galaxy rotation or even by stellar wind or stellar cycle, so the available gas mass for accretion increases, leading to an increase in the BH mass and the AGN luminosity. This period may vary from a galaxy to another depending on its size and morphology (there is a bar or not), where Alonso et al. (2014), Kim & Choi (2020) showed that the nuclear activity is higher in galaxies having a bar, referring to the vital role of the bar in transporting the gas to the central region. Generally, in this phase and through a transition for the total luminosity of the galaxy; the galactic disk luminosity may decrease gradually or becomes constant depending on the value of SFR and the AGN luminosity starts to increase. This phase can also be interpreted in terms of a study done by Zewdie et al. (2020) using SDSS MPA-JHU catalogue with the stellar mass range of $Log M_* = 10.73 - 11.03 \, M_\odot$. Using BPT-diagram, these authors found that the AGN, in this stellar mass range, have lower star formation rates than star-forming galaxies, and galaxies in this range move from the blue cloud to the red sequence.

**Figure 5.** The relation between the BH mass and the bolometric luminosity of AGN for the observed data of the galaxy sample (top panel). The black line is our results according to equation (12) and the blue line is the fit of observed data. The regular residual of equation (12) showing a normal distribution (bottom panel).

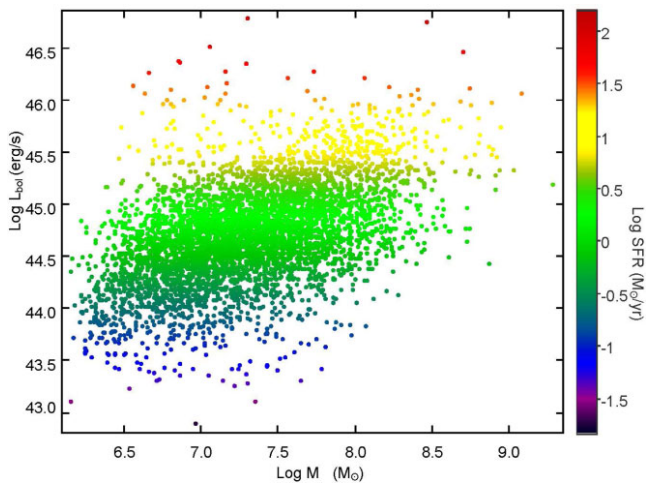

**Figure 6.** The BH mass and AGN luminosity against star formation rate distribution.

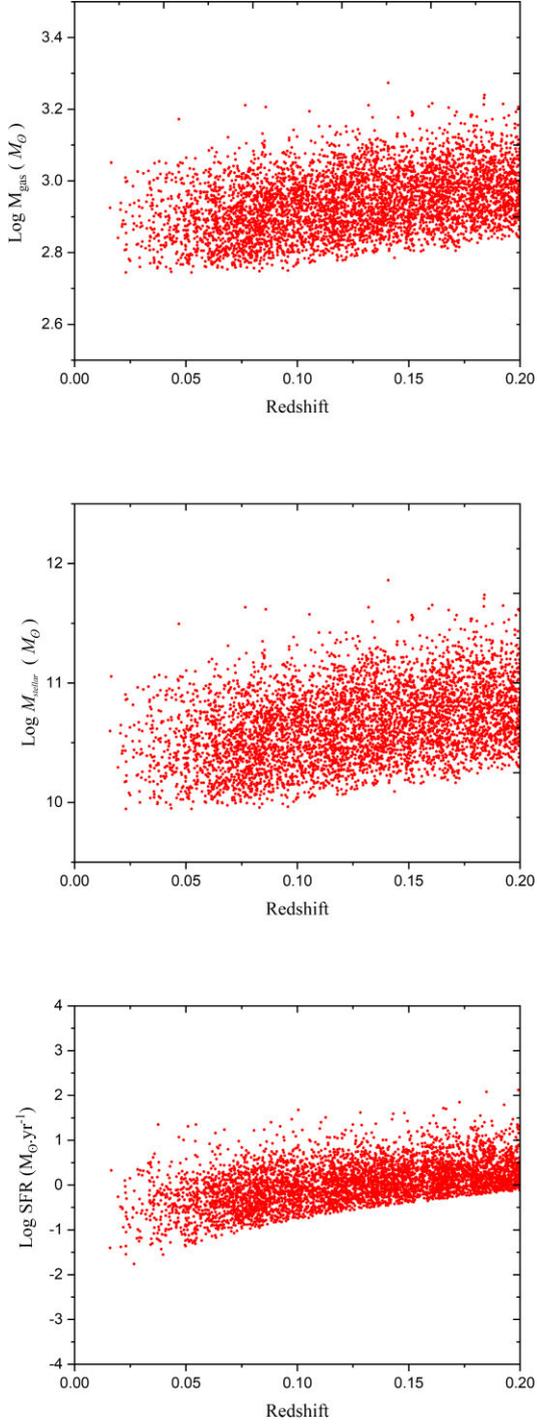

**Figure 8.** The gas mass (top panel), stellar mass (middle panel), and star formation rate (bottom panel) variation with redshift.

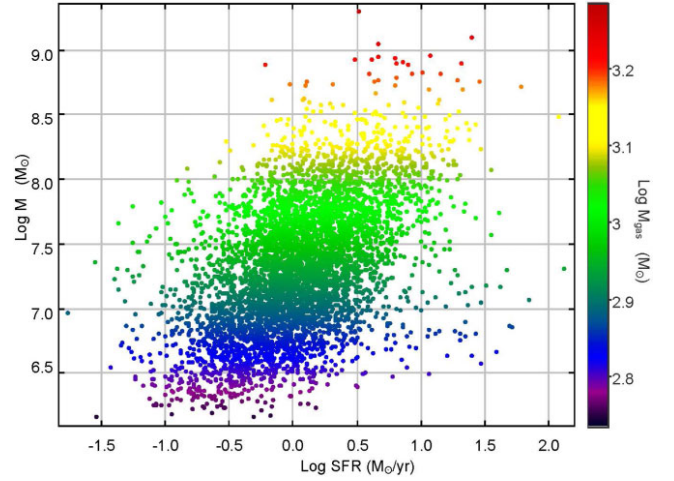

**Figure 9.** The relation between SFR, BH mass, and gas mass.

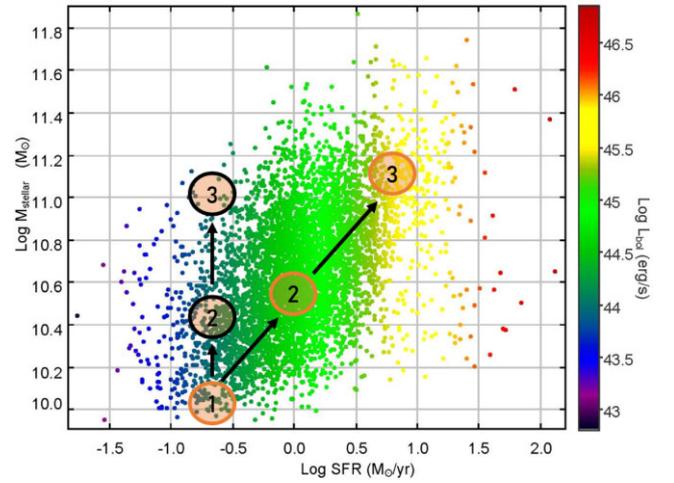

**Figure 10.** The 3 evolutionary phases of the spiral galaxy. In case of effect of AGN feedback (positive) on SFR (orange circles) and without the effect of AGN on SFR (black circles) feedback.

*Phase* 3 is the period of time where the accretion rate of central BH increases due to the large amount of gas coming from spiral arms. Accordingly, a small amount of gas remains in the spiral arms for star formation. So what we will observe is high stellar mass and low galactic disk luminosity but high AGN luminosity.

In Fig. 10, we summarize the 3 phases for the spiral galaxy evolution. If the AGN has no effect on the SFR of the host galaxy (black circles) therefore it has a nearly constant SFR, for example, for Log SFR $\sim -0.6 \, M_\odot \, yr^{-1}$. The stellar mass increases, leading to an increase in the galactic disk luminosity due to the stellar luminosities superposition but the increase in the AGN luminosity is very small, it changes slightly between Log $L_{bol} \sim 44$ and $\sim 44.5 \, erg \, s^{-1}$. In contrast, if the AGN has a positive feedback on the host galaxy, the SFR increases. A factor of 25 increase in SFR is related to an increase of 2 order of magnitudes in the AGN luminosity.

## 5 DISCUSSION

According to the results presented in the section 4, we find out that the torus properties such as spiral arm radius, spiral arm density and torus length can be used as indicators of the the evolution of the AGN. Also the accretion rate by the BH doesn't only depend on the BH physical parameters such as its spin, but also on the torus properties or the host galaxy properties. From our results we conclude that the AGN can have an effective role in the evolution of its host galaxy and vice versa. For each galaxy, the luminosity increases as the spiral arm decreases, which means that the gas in spiral arms is the main source for the AGN luminosity and changing it leads to a change in the AGN luminosity. Also accretion of stars onto supermassive BHs can occur. But to determine which is more efficient in the BH growth, the gas accretion or the star accretion, Pfister et al. (2021) treated the tidal disruption events caused by star accretion in cosmological simulations. These authors found that contribution from stars' accretion is negligible compared to gas accretion.

However this decrease in spiral arm gas content is not only due to the accretion by the BH but partially caused by the feeding process of BH and partially by the star formation taking place in spiral arms. The star formation rate is high when the gas mass is high in spiral arms, but to ease the gas flow from the spiral arms to the central BH, the gas should be dispersed and this is done by the consumption of gas in the star formation process and the dispersion caused by the stellar wind or any damping mechanism of the formed stars (Thompson, Quataert & Murray 2005; Hayward & Hopkins 2017; Lupi 2019).

The dispersion of gas which causes the gas to flow easily from the spiral arm toward the BH leads to a high AGN luminosity. This process is not continuous but occurs periodically depending on the star formation rate at spiral arms and the amount of gas available for it and it can be measured as a variability of AGN activity across the galaxy life time. This variability could make the galaxy normal for a period of time, and active for another period of time but, for further investigation of this variability, time-dependent SEDs are needed to be studied. Another discontinuity of this feeding process can be caused by the interaction between AGN feedback (outflows) and the material flowing from the spiral arm.

Due to the accretion process and feeding mechanism the spiral arm could disappear in a short period of time. But the accretion process is slowed down by the feedback coming from the AGN as its BH reaches a critical mass calculated by Ishibashi & Fabian (2012). And it is mentioned by Zubovas & King (2014b) that the BH reaches a critical mass in which the AGN begins to produce outflows and this may be one of the reasons for slowing down the accretion process. This leads to a slow increase in the AGN luminosity at higher BH masses and seen as a flattening in Fig. 5 (top panel). This slow gas consumption is also what might cause the spiral arm to be long lived.

All of the above results and discussion concern about an isolated galaxy without taking into account the environment effects such as merging of galaxies or the location of this galaxy in its cluster. Mergers and location of host galaxy with respect to cluster center can affect the time at which the gas of host galaxy disk is being consumed due to the AGN activity but not the physical process occurring between host galaxy disk and its BH. For example, wet merger can quick the process of accretion by triggering the gas into the BH. Also the motion of the galaxy can affect the life time of the process which can be slowed down due to the ram pressure causing loss of matter.

## 6 CONCLUSIONS

In this work, we have seen how the AGN can affect the evolution of the spiral galaxies and that our approach provides an evolutionary track for the AGN or specifically for the spiral galaxies in terms of their AGN evolution. This track begins with a BH of low mass, feeding on the gas mass of the host galaxy. Through its evolution and its consumption of gas mass, the luminosity increases then decreases slightly.

If we focus on this evolutionary track for AGN, we can see that as time evolves the luminosity decreases due to the decrease or the lack of gas mass in spiral arm which is also consistent with the decrease in spiral arm radius, and this results also were obtained by Masoura et al. (2018) for types 1 and 2 where the X-ray luminosity was found to decrease as the redshift decreases indicating a decrease in AGN activity.

From the observed data, the gas mass decreases with decreasing the redshift. Since the spiral arm density indicates the gas mass within the spiral arm, hence the spiral arm density also decreases with decreasing the redshift. So we can say that, during the evolution of AGN in spiral galaxies, the gas in spiral arms is consumed in feeding the central BH which indicates that the AGN is affecting the morphology of spiral galaxies. Hence we can link the AGN luminosity to the spiral arm radius or the gas mass in the spiral arm and use equation (11) to get the morphology of distant active galaxies through observing their luminosity and vice versa by assuming best fit values for each parameter that give the observed AGN luminosity.

This evolution has some consequence in between such as the variable appearance observed in AGN or the AGN variability. In studies done by Oh et al. (2015b) and Suh et al. (2015) for type 1 and type 2, the BH mass-luminosity relation was controlled by the Eddington ratio which indicates a change in the accretion rate of the central BH. This change can be explained in terms of our model approach by considering the spiral arm radius or the gas content which represents the gas reservoir for BH accretion rate. This also shows that the accretion process of the gas in spiral arms is not continuous but happens in phases or episodes of time. As mentioned by Zubovas & King (2014b) when the BH reaches a critical mass in which the AGN begins to produce outflows, this may be one of the causes for slowing down the accretion process. This slowing down is what causes the spiral arm to be long lived.

## ACKNOWLEDGEMENTS

## DATA AVAILABILITY

No new data were generated or analysed in support of this research

This paper has been typeset from a TEX/LATEX file prepared by the author.